# The Rashba Scale: Emergence of Band Anti-Crossing as a Design Principle for Materials with Large Rashba coefficient

Carlos Mera Acosta[1,2], Elton Ogoshi[2], Adalberto Fazzio[3], Gustavo Dalpian[2], Alex Zunger*[1]

[1]*Renewable and sustainable Energy Institute, University of Colorado, Boulder, CO 80309, USA*
[2]*Center for Natural and Human Sciences, Federal University of ABC, Santo Andre, SP, Brazil*
[3]*Brazilian Nanotechnology National Laboratory CNPEM, 13083-970, Campinas, SP, Brazil*

**Summary**

The spin-orbit -induced spin splitting of energy bands in low symmetry compounds (the Rashba Effect) has a long-standing relevance to spintronic applications and to the fundamental understanding of symmetry breaking in solids, yet the knowledge of what controls its magnitude in different materials is difficult to anticipate. Indeed, rare discoveries of compounds with large Rashba coefficients are invariably greeted as pleasant surprises. We advance the understanding of the 'Rashba Scale' using the 'inverse design' approach by formulating theoretically the relevant design principle and then identifying compounds that satisfy it. We show that the presence of energy band anti-crossing provides a causal design principle of compounds with large Rashba coefficients, leading to the identification via first-principles calculations of 34 rationally designed strong-Rashba compounds. Since topological insulators must have band anti crossing, this leads us to establish an interesting cross functionality of 'Topological Rashba Insulators' (TRI) that may provide a platform for the simultaneous control of spin splitting and spin-polarization.

# INTRODUCTION

Spintronics aims at generation, detection, and control of the spin degrees of freedom with the relevant functionalities being generally based on the magneto-electrical generation of spin-polarized states[1–3]. E. Rashba noted in 1959 [4,5] that when an asymmetric electric potential breaks inversion symmetry, spin-orbit coupling (SOC) creates an effective magnetic field that leads to spin-split and polarized bands. The magnitude of the effect[6,7] is given by the ratio between the spin splitting $E_R$ and the momentum offset $k_R$, that is $\alpha_R = 2E_R/k_R$. Strong and weak Rashba effects are defined by the measured or DFT calculated magnitude of the Rashba coefficient. Symmetry wise, the existence of a Rashba effect of arbitrary magnitude $\alpha_R$ requires a compound with non-centrosymmetric structures having local electric dipoles induced by polar atomic sites that add up over the unit cell to a non-zero [8,9]. Despite the fact that large Rashba effect is needed for facile spintronic generation and detection of spin-polarized states[10–13] as well as for the detection of Majorana Fermions[14,15], the principles determining the *magnitude* of this functionality ('the Rashba scale') has not been established. Indeed, the discovery of new compounds with large Rashba coefficient (e.g., GeTe (*R3m*)[16,17], BiTeI (*P3m1*)[18,19], and metallic PtBi$_2$ (*P3m1*)[20]) is invariably greeted as a pleasant surprise. The few available literature calculations of compound with significant $\alpha_R$ and the general absence of examples of weak Rashba effect compounds poses a severe bottleneck to the understanding of the underlying physical factors controlling the trends, as well as to the prospects of advancing effective spintronic technology.

We show in this paper that the magnitude of Rashba coefficients in different compounds is not well correlated with the magnitude of the spin orbit coupling, and that the hallmark of strong Rashba coefficient is the appearance energy band anti-crossing of the Rashba split bands.    This has a few immediate consequences: First, because all topological insulators must have band anti crossing, we show that that *all* non-centrosymmetric topological insulators having non-zero electric dipole (i.e., TI that can have a Rashba effect) must be *strong* Rashba compounds. This provides a causal physical explanation for previous occasional observations of TI's having large Rashba coefficients[21–24], and establishes a new *cross-functionality: Topological Rashba Insulators (TRI's).* Searching current data bases of TI compounds (Refs [25], [26], and [27]) for TI members that are also non-centrosymmetric with non-zero electric dipole predicts a few *TRIs* with such as Sb$_2$Te$_2$Se and TlN with calculated large $\alpha_R$ of 3.88 eVÅ and 2.64 eVÅ in the valence bands, respectively. Second, we show that the anti-crossing theory of the Rashba scale can be used to identify new strong Rashba compounds by a different route -- starting from known non-centrosymmetric structures and identify those that also have anti-crossing bands. This approach led to identification of 34 previously synthesized strong Rashba compounds, including the already known GeTe and BiTeI, as well as compounds that have been previously synthesized, but were unappreciated as Rashba compounds, let alone as strong Rashba compounds such as BiTeCl (*P6$_3$mc*), PbS (*R3m*), and K$_2$BaCdSb$_2$ (*Pmc2$_1$*) with Rashba coefficients of 4.5, 4.6, and 5.3 eVÅ, respectively. Additionally, we also identify 165 weak Rashba compounds with Rashba parameter smaller than 1.2 eVÅ and RSS large than 1 meV. We hope that these predictions will be tested experimentally.

The theory above follows an *Inverse Design* approach: It predicts target properties based on physically-motivated models that directly connect the existence of the desired property with an explicit physical mechanism[28–31]. Searching of specific realizations of such materials is then performed by first principles calculations, looking for the above-established metric of the physical mechanism in real materials. This is different than an exhaustive search data-directed approach in which the discovery of materials with a given functionality is based on high-throughput computation of all (or many) possible combinations of atomic identities, composition and structures[32,33]. This is also different from traditional machine learning, in that inverse design relies on the use of an explicitly causal physical mechanism rather than on the correlation of say, atomistic features with the target functionality[34–36].

The main accomplishments of the current work are: *i*) the development of the definition of the Rashba scale: all materials with larger than certain value $\alpha_R$ have band anti crossing, and below that threshold none has band anti-crossing, *ii*) the demonstration of how anti-crossing bands can be identified form the atomic orbital contribution to the band structure, *iii*) the establishment of TRIs, *iv*) the inverse design of 34 strong Rashba compounds and 165 weak Rashba compounds based on the proposed theory, i.e., the anti-crossing as design principle for strong Rashba materials. The advance offered by this establishment of a bridge between electronic structure (viz. band anti crossing) and the "Rashba scale" may offer a platform for the exploration of other phenomena potentially hosted by Rashba compounds, e.g., superconductivity and Majorana Fermions.

## RESULTS

### Shortcomings in the current understanding of trends in the Rashba scale

To discuss trends in the Rashba scale, Figure 1a presents DFT-calculated Rashba coefficient (See Methods section for details of calculations) of 125 compounds that have larger than 1 meV spin splitting near the valence band maximum. This gives a broader impression of the distribution of the magnitude of the Rashba coefficients than what is currently available from isolated literature calculations. We focus on compounds with intrinsic dipoles, ("Bulk Rashba effect" [13], denoted as R-1). We exclude (i) magnetic compounds (no time-reversal-symmetry) in which the Zeeman effect is observed instead[37,38], (ii) surfaces/interfaces-induced Rashba effects (the "R-0" effect)[7,39], which require non-bulk symmetry breaking, and (iii) centrosymmetric compounds with local sectors that have non centrosymmetric point groups ('hidden Rashba effect' [8,9] or "R-2"). Fig. 1a shows the existence of a significant range of $\alpha_R$ and a general delineation (marked approximately by the blue hatched lines) into small vs large band edge Rashba effects, which are based on $\alpha_R$ *and* hereafter referred to as weak vs. strong Rashba, respectively.

In the phenomenological Hamiltonian describing the linear-in-*k* Rashba effect in quasi two-dimensional systems[6,7],

$$H(k) = -\sigma_0 \frac{\hbar^2 k^2}{2m^*} + \alpha_R(\sigma_x k_y - \sigma_y k_x), \qquad (1)$$

the magnitude of the Rashba coefficient $\alpha_R$ is associated with the intrinsic atomic SOC. However, the continuum $k \cdot p$ theory underlying the literature based on Eq. (1) does not disclose trends in the magnitude of $\alpha_R$, for which an atomistic resolution is needed. Indeed we will show that *all materials with larger than certain value $\alpha_R$ have band anti crossing, and below that threshold none has band anti-crossing. Furthermore,* the *continuum-like Rashba Hamiltonian of Eq(1) mimic microscopic energy level quantum models only for the specific cases of non-crossing bands.*

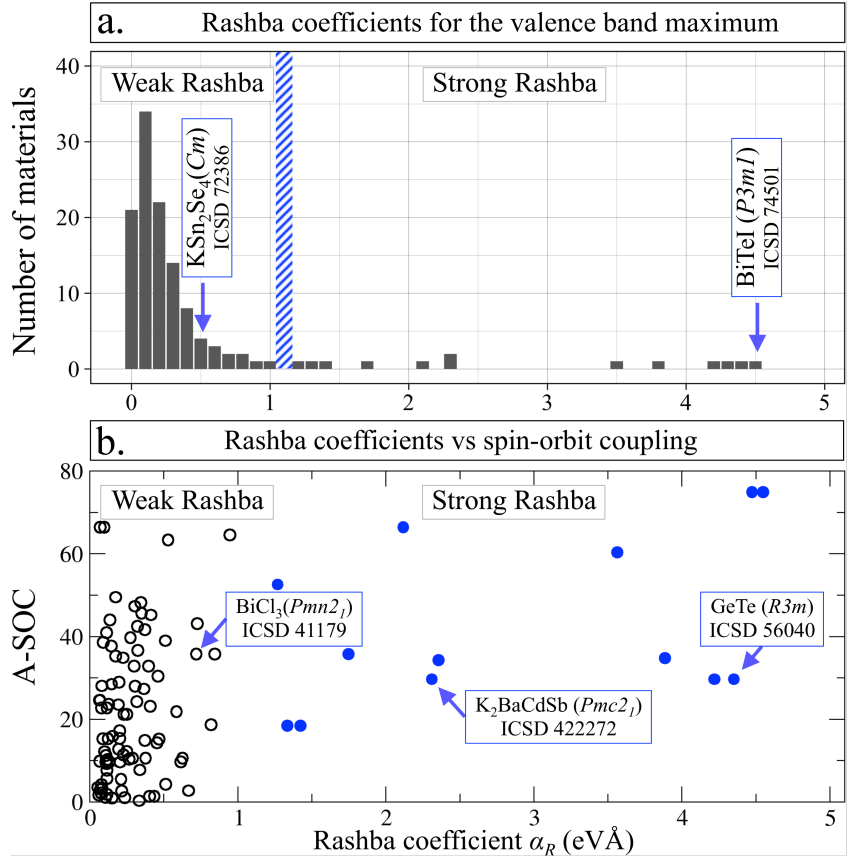

**Figure 1.** (a) DFT calculated (see Method section for details) Rashba coefficients for the valence band maximum of 125 compounds. The plot reveals a general delineation into "strong" (illustrated by BiTeI) and "weak" (illustrated by KSn$_2$Se$_4$) Rashba coefficients. Their Inorganic Crystal Structure Database (ICSD) code and space group are indicated in Table II and supplementary information II, respectively. The blue-hatched area indicates the general delineation. (b) Rashba coefficients vs. an average of the atomic spin-orbit coupling weighted by the composition (A-SOC) for the valence band maximum of 125 compounds, for weak (unfilled black dots) and strong (filled blue dots) Rashba compounds.

In 3D compounds, the Bulk Rashba effect can depends on the interatomic orbital interaction, hindering the description of this effect by Eq. (1) even in planes perpendicular to the electric dipole. Inspection of DFT results in Fig 1 shows, however, that this scaling through SOC is not the whole picture. Figure 1b shows the Rashba coefficient at the VBM plotted vs. the composition-weighted average of the atomic SOC values (A-SOC) of the respective compounds (taken from Ref [40]). This reveals that for compounds defined as 'weak Rashba' (open black circles) there is a generally non-monotonic trend of $\alpha_R$ with A-SOC, making it unlikely to predict a sequence of compounds with monotonic $\alpha_R$ values based on A-SOC alone. Thus, compounds with lower SOC can have larger Rashba coefficients than those with higher SOC. This is illustrated for instance by K$_2$BaCdSb$_2$ (space group *Pmc21*) having large calculated $\alpha_R$ of 2.36 eVÅ for the valence band and 5.25 eVÅ for the conduction band, while having a *smaller* atomic A-SOC than BiCl3 (space group *Pmn21*) with $\alpha_R = 0.72$ eVÅ for the valence band and $\alpha_R = 0.403$ eVÅ for the conduction band.

**Role of orbital interactions and band shapes in determining the Rashba Scale**

The definition of $\alpha_R = 2E_R/k_R$ suggests that a large Rashba coefficient must be a statement of large energy splitting $E_R$ obtained in a short momentum step $k_R$, whereas small Rashba coefficient necessarily means small energy split achieved in a long wavevector step. Such different dispersion curves are indeed apparent in previous DFT calculations as illustrated in Fig. 2 for the prototypical band shape in BiTeI and KSn$_2$Se$_4$ with Rashba coefficient in the VBM of 4.6 and 0.6 eVÅ, respectively. One notices qualitatively different behaviors of the dispersion shape of the Rashba bands of strong Rashba compounds vs. weak Rashba compounds: BiTeI (space group *P3m1*) has a significant bowing of the bands with small momentum offset and large RSS, compare to KSn$_2$Se$_4$ (space group *Cm*) with its weakly dispersed band, large momentum offset, and small RSS. These trends translate into large $\alpha_R$ (as in BiTeI) and small $\alpha_R$ (e.g., KSn$_2$Se$_4$).

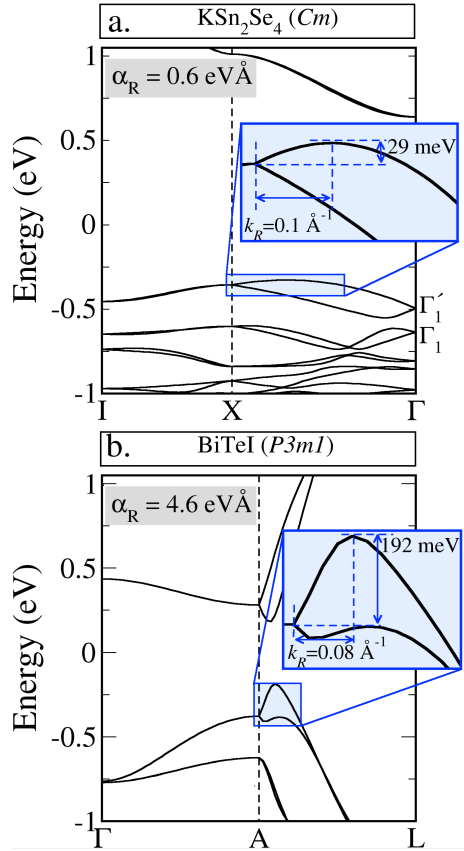

**Figure 2.** DFT band structure calculations of prototype compounds characterized by different band shapes: (a) without anti-crossing bands, and (b) with anti-crossing. This results in small ($\alpha_R$=0.6 eVÅ) and large ($\alpha_R$=4.6 eVÅ) Rashba parameters, respectively. The different band shapes, momentum offset and Rashba spin splitting are highlighted in the blue insets. In KSn$_2$Se$_4$, the valence bands have the same symmetry representation ($\mathbf{\Gamma_1}$ and $\mathbf{\Gamma_1'}$).

**Tight Binding model that allows continuous transition between weak and strong Rashba behaviour**

To establish if these characteristic dispersion shape bear a causal relationship to the magnitude of the Rashba coefficient also *in a* given material (rather than in *chemically dissimilar compounds* as KSn$_2$Se$_4$ and BiTeI), it would be useful to control the dispersion *shapes in the same material*. Alas, this is not easy to do with DFT since the band shape can strongly depend on the atomic composition, lattice symmetry, and specific orbital interactions. But such shape engineering of band dispersion is readily possible within a tight binding (TB) model that, however, does not have the additional virtue of material realism. Our strategy is therefore to use a simple TB model that enables transmuting the shape of Rashba band dispersion between the two prototypes of Fig 2, thus establishing what controls large vs. small Rashba effects in a toy model, then use this TB identification *of a metric* in precise and material specific (3D) DFT calculations and observe how this reveals strong vs weak Rashba effects in real compounds. To this end, we constructed a model Hamiltonian including the minimal essential ingredients at play, namely: two orbitals at different sites, opposite effective mass sign, and SOC ($t_{soc}$). For illustrative purposes, we only consider *s*- and *p$_x$*-orbitals interacting through the hopping term $t_{sp}$. A detailed description of the effective tight-binding Hamiltonian used here is given in the Methods section.

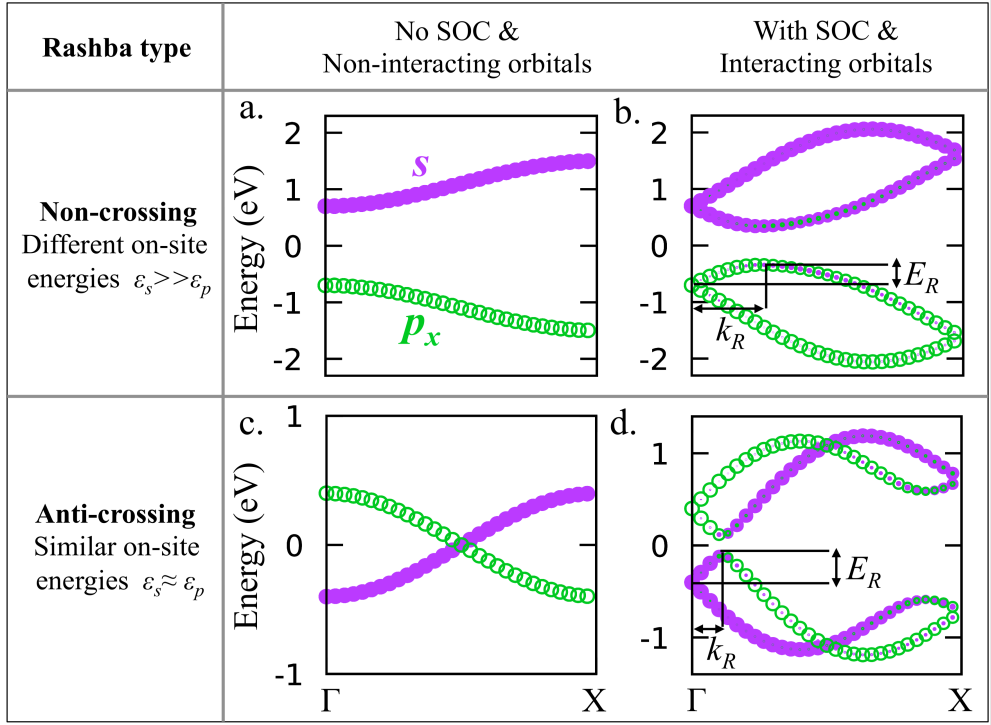

**Figure 3.** Evolution of the band structure along the line Γ-X for both kinds of Rashba (non- crossing bands ($\varepsilon_s=-\varepsilon_p=0.55$ eV) and with anti-crossing bands ($\varepsilon_s=\varepsilon_p=0$ eV)) for the case of: (a) and (c) no SOC and no inter-orbital coupling *sp* ($t_{sp}$); and (b) and (d) SOC ($t_{soc}$) and inter-orbital interaction. Here, $t_{soc}$=0.4eV, $t_{sp}$=0.3eV, and $t_{pp}$=-$t_{ss}$=0.2eV. For crossing bands, the *sp* interaction leads to band interaction, which in turns causes anti-crossing bands, meanwhile non-crossing bands are weakly affected. When the SOC is turned on, bands with and without crossing respectively lead to small (b) and large (d) Rashba coefficients, giving the differentiation of the band dispersion of weak and strong Rashba effects.

For the TB parameter set corresponding to no SOC and non-interacting bands, we find, as expected, non-crossing bands (i.e., different on-site energies $\varepsilon_s>\varepsilon_p$) having parabolic shapes with no Rashba effect (Fig. 3a). When these bands are allowed to interact (via setting $t_{sp}>0$) and experience SOC, as in Fig 3b, the emerging Rashba band shapes is typically "small $E_R$ and large $k_R$", with its small attendant Rashba coefficient. To change qualitatively the dispersion shape to "large $E_R$ and small $k_R$" one needs to bring the non-interacting bands (shown in Fig. 3c) closer to each other, (e.g., by making the on site energies similar $\varepsilon_s \approx \varepsilon_p$). Notably, when the non-interacting crossing parabolic bands (Fig. 3c) are allowed to experience SOC and interact (Fig. 3d), this orbital interaction (of the same magnitude as in Fig. 3b) leads to *band anti-crossing* with large and linear Rashba effect. This also provides a qualitative description of the typical orbital character behaviour in band anti-crossing, i.e., the orbital character drastically changes (e.g., form *s*- to $p_x$-orbitals in the VBM shown in Fig. 3d) as the *k*-vector changes from smaller to larger values than $k_R$ (Fig. 3d). The predicted band shapes in the linear Rashba effect with non-crossing bands (Fig. 3b) and with anti-crossing bands (Fig. 3d) provide a differentiation between the band dispersion of weak Rashba compound illustrated by DFT calculation on KSn$_2$Se$_4$ (Fig. 2a) vs. strong Rashba compound illustrated by DFT calculation on BiTeI (Fig. 2b). *Thus, band anti crossing due to band interaction is the deciding factor, within the simple TB model, for the transition between the weak Rashba to the strong Rashba band shape behavior.*

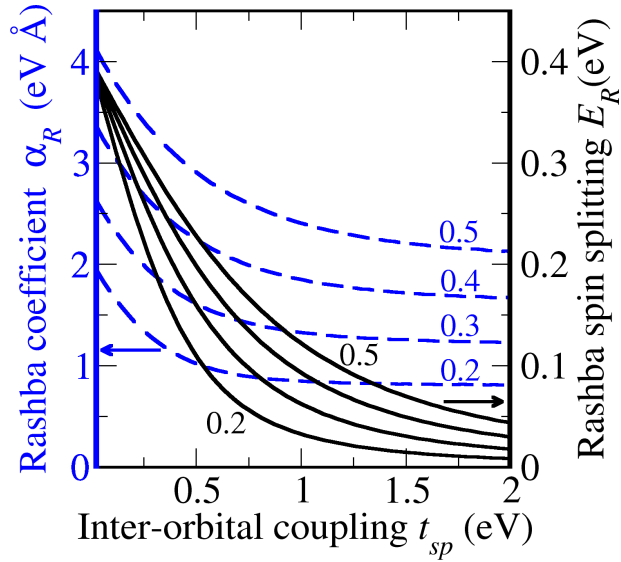

**Figure 4.** Variation of the Rashba coefficient $\alpha_R$ (dashed blue) and Rashba spin splitting (solid black) as a function of s-p orbital coupling for anti-crossing bands. Different values of the SOC are denoted for each blue line. Note that the variation of the RSS as a function of the band interaction is slower as the SOC increase.

The TB model provides insight to the behavior of the classic Rashba Hamiltonian Eq. 1. *This classic Rashba Hamiltonian mimics TB only for the specific cases of non-crossing bands*: in TB, for non-interacting bands, the diagonal elements of the block diagonal Hamiltonian describe single-orbital bands, leading to the expression $H_p(k) \approx -\sigma_0(t_{pp}a^2 k_x^2) + 2at_{soc}^p(\sigma_y k_x)$ for *p*-orbitals (with $t_{pp} = \varepsilon_p/2$), as shown in Methods section. Thus, $H_p(k)$ is equivalent to Eq. 1 (by taking $t_{pp}a^2 = \hbar^2/2m^*$ and $\alpha_R = 2at_{soc}^p$ for $k_y = 0$). This illustrates that for *non-crossing bands*, $\alpha_R$ is proportional to the SOC (i.e., $\alpha_R = 2at_{soc}$), and decrease as the inter-orbital interaction increases since orbitals are deformed by the atomic bonding[41,42]. However, for *anti-crossing bands in TB*, $\alpha_R$ and RSS depend on the inter-orbital coupling strength (an effect absent from Eq (1)), as shown in Fig. 4. We see that with *weak inter-orbital interaction*, $\alpha_R$ is much larger *than the SOC itself* ($\alpha_R = 4$ eVÅ for a SOC of 0.5 eV) and the RSS also reach large values (here, 400 meV) (Fig. 4). Both $\alpha_R$ and RSS decrease monotonically as the orbital interaction increases. For strong inter-orbital interaction, the $\alpha_R$ reaches a constant value corresponding with the Rashba effect without anti-crossing. The RSS tends to values smaller than 50 meV, showing that even for anti-crossing bands, while the Rashba parameter is large, the RSS is not necessarily large.

### DFT validation of the role of band anti crossing in the Rashba scale

As noted before, the TB model lacks material realism. But we can test via realistic DFT calculations the central insight it provides: We can directly detect in 3D DFT calculations with SOC which compound has band anti crossing and distinguish it from compounds that lack band crossing. This is done by orbital-projected band structure, i.e., calculating the weight of the atomic orbitals in the wave function for each *k*-point and each band index. That is, the qualitative different band shapes in TB depiction of compounds with large $\alpha_R$ (Fig. 3d) and small $\alpha_R$ (Fig. 3b) are linked to the realistic DFT depiction via the atomic orbital contributions to the band structures where band anti-crossing is directly identified by verifying the existence of orbital character change as the momentum goes from $k < k_R$ to $k > k_R$. We have studied the band crossing vs. band anti crossing behavior of the compounds shown in the survey Fig 1.

We show In Figure 5 the DFT-calculated band shape and orbital-projected band structure predicted in Fig 1 to be strong Rashba compounds. For instance, in BiTeI (Fig. 5a), for $k < k_R$, the Te-$sp_z$ (Te-$p_{xy}$) orbital contributes to the VMB (CBM), but for $k < k_R$, this orbital contribution moves to the CMB (VBM), as indicated by the magenta (green) dashed line. This indicates the existence of band anti-crossing. *All DFT confirmed strong Rashba compounds clearly show band anti-crossing* (see orbital-projections for PbS, Sb2Se2Te, and GeTe in Figs. 5b-d). Thus, the DFT-calculations are in agreement with the physical causal relation between the existence of anti-crossing bands and strong Rashba effect. This definition of the Rashba scale also provides a numerical description of the strong Rashba effect, i.e., all compounds with Rashba coefficient approximately larger than 1.3 and 1.6 eVÅ in the VBM CMB (See Table III), respectively, are also strong Rashba compounds. We term compounds with strong Rashba (i.e., large $\alpha_R$ and band anti-crossing) as Type I, whereas compound with weak Rashba (i.e., small $\alpha_R$ and no anti-crossing) as Type II.

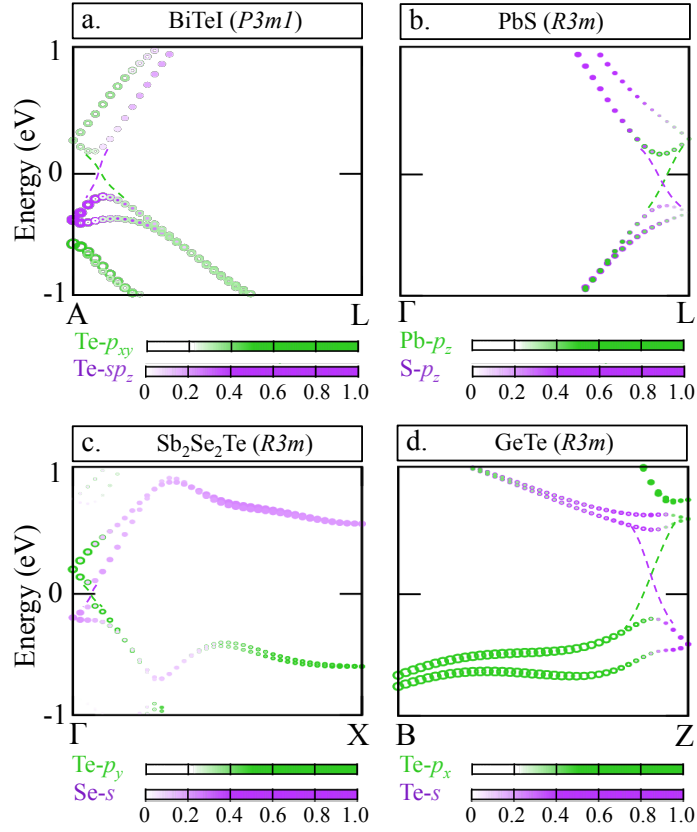

**Figure 5.** Orbital-projected band structure for the strong Rashba materials (a) BiTeI, (b) PbS, (c) $Sb_2Se_2Te$, and (d) GeTe (codes 74501, 183243, 60963, and 56040 in the inorganic crystal structure database[43], respectively). The green and magenta color scales stand for the orbital contribution to the CBM and VBM at the time-reversal high symmetry $k$-points, respectively. These orbital contributions change for momentums larger than the momentum offset, indicating the existence of anti-crossing bands. Only orbitals defining the anti-crossing are shown. The size of the dots also varies according to atomic orbital weight in the wavefunction of each k-point and band index. Dashed green and magenta lines are shown to guide the view to the change of the orbital character.

### The emerging Cross-Functionality of 'Topological Rashba Insulators' (TRI)

We next explore some of the consequences of the above-noted definition of the Rashba scale i.e. that a *Type I* Rashba compound can only be found in compounds featuring also energy band anti-crossing.

There is a class of material functionality that is characterized by always having energy band anti crossing, namely topological insulators (TI)[44]. TIs have an *inversion* in order between valence and conduction bands; however, this does not guarantee the energy band anti-crossing (e.g. HgTe has inversion in band order even when calculated without SOC[45]). Band *anti-crossing in TI's* is only created if the interaction of the inverted bands is symmetry allowed[46], whereas when the interaction between inverted energy bands is symmetry forbidden, the compound exhibit is a topological metals, not insulators[47]. Given that TI always has band anti crossing, and that strong Rashba compounds must have band anti crossing, we next enquire what are the additional conditions for a TI to have a Rashba spin splitting (so it would be a *Type I* material).

We recall that the symmetry condition the Rashba R-1 effect[8] is that the compound must be non-centrosymmetric with a non-zero local electric dipole that add up over the unit cell to non-zero value. Thus, according to the proposed theory, *all non-centrosymmetric TIs having local dipoles that add up to non-zero are strong Rashba compounds*. This observation will be used below to explain previously puzzling observation of trends in Rashba effects in TI's, and to identify compounds that have the cross-functional property of TI's while also being Rashba materials (TRI's).

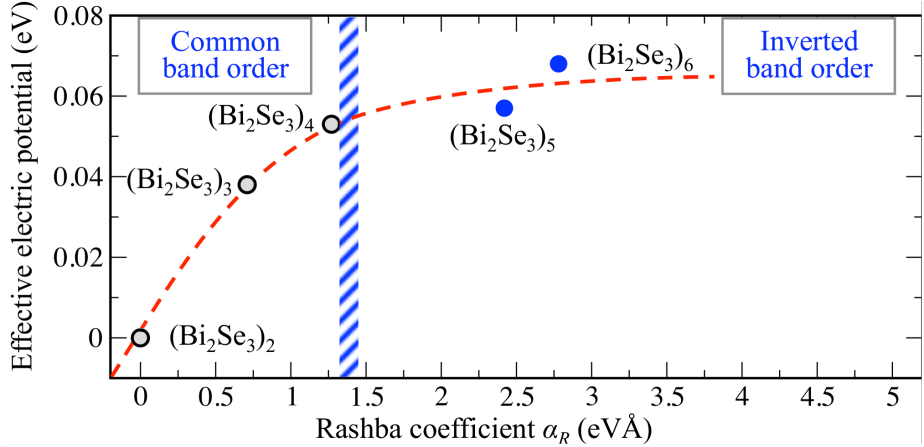

**Figure 6.** Variation of the experimentally estimated Rashba coefficient $\alpha_R$ and effective electric potential as a function of the slab thickness in the topological insulator $Bi_2Se_3$[21]. Slabs with common band order (non-topological insulators) and inverted band order are represented by the gray and blue dots, respectively. An abrupt change in the Rashba coefficient is observed when the band order changes. The blue-hatched area (indicated also in Fig. 1) separates slabs with different band order and small from large Rashba coefficients. The red shows the trends of the Rashba coefficient as function of the effective electric dipole.

Experimental evidence of trends in Rashba behaviour in non-cetrosymmetric TIs was observed in thin films of n formula units $(Bi_2Se_3)_n$ of the TI $Bi_2Se_3$ grown in a SiC substrate[21]. Figure 6 shows the experimentally estimated Rashba coefficient for different repeat units n in $(Bi_2Se_3)$ plotted against the estimated effective electric potentials that reflects the breaking of inversion symmetry (generated here by the induced electric dipole of SiC substrate). As seen in Fig. 6, this electric potential changes when n increases, but it remains almost the same for n=4 and 5 so both the SOC and the electric dipole are almost constant for these n values. This leaves unexplained the $\Delta\alpha_R = 1.15$ eVÅ jump in the Rashba coefficient between with n=4 and n=5 despite having the same SOC and potential asymmetry. This surprising fact is, however, in agreement with the band anti crossing theory of the strong Rashba effect as $Bi_2Se_3$ thin films as band inversion (and band anti-crossing) have been predicted to take place only for $(Bi_2Se_3)_n$ with n>4[21], as indicated in Fig. 6. In other words, the inversion in the band structure is accompanied by an abrupt change in the Rashba coefficient $\Delta\alpha_R$.

### The way to identify topological insulator compounds that are at the same time Rashba materials (TRI's)

Finding cross functionalities is always interesting, such as multiferroics[48,49], Ferroelectrics that are Rashba[50], transparent conducting compounds[51–53] and electrical conductors that are thermal insulators [54,55]. Topological Rashba insulators will have spin split surface states, an interesting, yet unobserved behavior. The task of identifying compounds that are TIs and Rashba starts by finding TI's (steps a-c in Table I) and then filtering out those TI that have at least one polar atomic site in the unit cell, and a non zero total dipole (steps d-e), i.e. qualify as Rashba. According to the foregoing band anti crossing theory, Rashba compounds that are TI must be **strong** Rashba compounds.

**A. Finding TI's**

(a) *Finding compounds that have their band structure computed by DFT +SOC:* We use literature databases of Refs [25], [26], and [27] that were obtained by screening the Inorganic Crystal Structure Database[43] (ICSD), including now a total of **203,380** entries, However, to determine TI-ness of a compound one needs[25] to compute its band structure including SOC. This requirement has drastically reduced the fraction of 203,000 ICSD compounds simply because for ~90 % of ICSD compounds the calculation of the band structure was deemed problematic for one reason or another. The reasons (theoretical, computational, structural, financial are different for Refs [25], [26], and [27]) are summarized in Supplementary information III. The results of these initial restrictions is (line (a) in Table I) that Ref [25] inspected **22,652** compounds as TI candidates, Ref [26] inspected **19,143** compounds, while Ref [27] inspected only **13,628** compounds (see Table I line a).

(b) *Find the fraction of compounds that can be symmetry protected topological phases (metals or insulators):* Given these restricted lists of potential TIs, the literature has then applied filters guaranteeing compounds with symmetry indicators of topological phases, i.e., capable of having an inversion in the order of bands. This is based on the topological class defined in terms of "elementary band representations"[44], symmetry indicator[56], or topological invariant[46]. This filter leaves **7385, 1075, and 4050** topological materials (either metals or non metals) taken from Refs [25], [26] and [27] respectively (line b in Table I).

Table I. Screening of topological Rashba materials. The first filter is the initial restrictions that the repositories applied to the ICSD to find a shorter list for which calculations have been done. Subsequently, the applied filters select all TIs from these shorter lists; all compounds with band gap larger than $10^{-4}$ eV (non-zero band gap); compounds with at least one polar atomic site (i.e., polar space groups); and finally, compounds with non-zero total dipole.

| Filters | Ref. [25] | Ref. [26] | Ref. [27] |
|---|---|---|---|
| a. Shorter lists obtained from ICSD | 22652 | 19143 | 13628 |
| b. Symmetry protected topological phases | 7385 | 1075 | 4050 |
| c. Non-zero band gap | 277 | 273 | 50 |
| d. At least one polar atomic site | 15 | 18 | 7 |
| e. Non-zero dipole | 0 | 0 | 3 |

(c) *Find the fraction of topological compounds that are topological insulators:* We then select compounds reported as non-metals (band gaps ($E_g$) larger than $10^{-4}$ eV), which results in a considerable reduction of the databases, i.e., **277**, **273** and **50** topological insulators, respectively **(line c in Table I).** We note that such tiny band gaps hardly qualify as "insulators" (despite the ubiquitous use of that term instead of "non metals" to describe arbitrarily small band gaps). For example, inspection of the 277 nonmetallic topological compounds of Ref [25] for those with a DFT gap of at least 0.1 eV or 0.5 eV leaves 34 and 0 topological narrow-gap semiconductors, respectively. The condition of non-zero band gap **(line c), which** is not related to the Rashba effect but required to guarantee anti-crossing bands, leads to a abrupt decrease in the yield of qualifying compounds. Interesting observations are that topological insulators are rather rare among the ICSD compounds examined (far more than, e.g., superconductors) and the vast majority of topological phases found are metals, being of less interest for physics that occurs inside the band gap, such as transport, Rashba effect, and topological surface states.

**B. Find TI's that are allowed Rashba compounds**

This entails two steps: Having used the literature selection of inorganic topological TIs, we finally consider the fraction that qualify as potential Rashba R-1 compounds, i.e., have at least one polar atomic site and a non-centrosymmetric space group.

*(d) Find the fraction of TI's that are non-centrosymmetry:* To this end we select out of the compounds that are topological non-metals (step c) compounds with space groups having at least one polar atomic-site. The list of point groups with at least one polar site is given in Figure S1 of supplementary information IV. This leaves us in step d with 15, **18** and **7** compounds from lists of Refs [25], [26], and [27] (line d in Table I), respectively. We note that in such compounds with at least one polar atomic site the necessary electric dipole for the Rashba effect can still be zero. Thus, step (e) is needed:

*(e) Find the fraction with* non-zero total dipole: In order to guaranteed Rashba-ness, the last applied filter distills compounds with non-zero total dipole (**line e in Table I**). The existence of finite net dipole is determined by local asymmetric charge distributions that add up to nonzero. These local charges are induced by interatomic bonding, which can be distributed in such a way that the dipole vectors generated by each bonding accidentally cancel each other (See supplementary information IV). The three TI databases of Refs [25], [26], and [27] leave only **0, 0, and 3** cross-functional TRI compounds, respectively: **$Sb_2TeSe_2$, $K_5Fe_2O_6$, TlN**. Unfortunately, according to our own DFT band calculation, $K_5Fe_2O_6$ is more stable in a ferromagnetic configuration (with $E_{AFM}-E_{FM}$=2.7 eV/per formula), meaning that the time reversal-symmetry is not preserved, so its not an R-1 compound. We tested our method of deduction by calculating in DFT the band anti crossing and Rashba coefficient of **$Sb_2TeSe_2$ and TlN** in Fig 7.

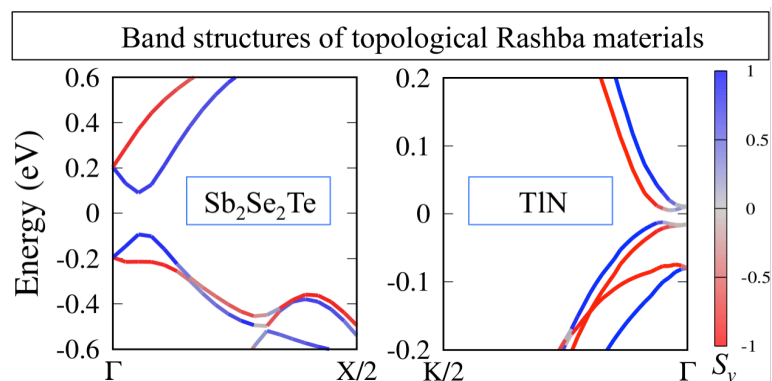

**Figure 7.** Band structured of the predicted topological Rashba $Sb_2TeSe_2$ and TlN. The color scale stands for the expected values of the spin operator $S_y$.

DFT calculations for the screened compounds (Fig. 7) verify that these are correctly predicted as strong Rashba semiconductors, as we discus below. The two TI compounds

Sb$_2$TeSe$_2$ (*R3m*)[57] and TlN (*P6$_3$mc*)[58] have been synthesized and predicted by our calculation to have a rather large Rashba parameter of 3.88 and 2.6 eVÅ for Sb$_2$TeSe$_2$ and TlN, respectively. The band structures of these compounds are shown in Fig. 7. Both compounds are classified in Ref. [27] as topological insulators protected by the TR-symmetry. Our calculated DFT band gaps are 179 meV and 18 meV for Sb$_2$TeSe$_2$ and TlN, respectively. This suggests that TlN is near a topological transition with a small spin splitting of 6 meV. The robust TIs Sb$_2$TeSe$_2$ has a very large spin splitting of 166 meV.

The interesting albeit disappointing result is that as we start from extensive lists of thousands of symmetry protected topological materials and then impose conditions for Rashba-ness we find only 2 strong Rashba compounds (**Sb$_2$TeSe$_2$, TlN**). This very small yield might suggest that perhaps "TI-ness" and "Rashba-ness" might be somehow contraindicated. But more likely is that the currently available list of TI with good insulating gap whose band structure has been calculated is very small: if a broader list of TIs compounds would be available (step a and b in Table I), more Rashba compounds with large coefficients might be identified: Note that the initial restrictions in step (a) to <10 % of the known inorganic compounds could unfortunately exclude some important Rashba candidates (See Supplementary information III). Furthermore, the condition of non-zero band gap **(line c in Table I),** leads to small yield of only 2% (<277) of inorganic compounds that are TIs, and even fewer (34 compounds) if the minimum gap has to be 0.1 eV.

Most importantly, considering the condition of **non-cetrosymmetric** TIs, the fraction is less than 0.1% of the initial shorted lists (e.g., only 15 NC-TIs in the list of 22652 compounds of Refs [25]). This means that the number of **NC-** TIs is small in the reporter lists, which is not related to existence of Rashba materials in nature. Indeed, we emphasize that these filters (band gap and NC space groups – lines c and d in Table I) are not conditions for the specific selection of either weak or strong Rashba compounds. The highlighted here is that all selected TRIs are predicted to be strong Rashba compounds, as predicted by the proposed definition of the Rashba scale as a consequence of the existence of energy bands anti-crossing.

**Discovery of strong Rashba compounds via DFT prediction of band anti-crossing**
The study of the previous section looked for the interesting cross functionality of topological insulators that also Rashba compounds, *starting from TIs* and down selecting those that are Rashba like. The complimentary search ignored thus far, starts from Rashba compounds and down selects those that have anti crossing bands even if they are not TIs. It turns out that the yield of this complimentary search is much larger than the previous search.

As we will start from Rashba compounds, one needs to note that there are a few types of Rashba band splitting compounds: either when the splitting is between different valence bands, or between different conduction bands, or between valence and conduction bands. For instance, as Fig. 2a shows for KSn$_2$Se$_4$, the interaction between the valence bands $\Gamma_{1v}$ and $\Gamma'_{1v}$ along the $\Gamma$-X symmetry path is symmetry allowed, leading then to a strong Rashba effect *inside the valence bands*. However, there is no anti-crossing between valence and conduction bands, so the Rashba effect at such *band edge* is weak. Here, we are interested primarily in compounds featuring *band edge Rashba splitting*, i.e., near the VBM or CBM. To this end we will focus only on anti-crossing between these band edge states.

Figure 8 describes the selection strategy based on our design principles, which is divided into 3 filtering process, shown in the vertical column in Fig. 8. The supplementary information IV provides more technical details on the selection strategies. We consider a database of Rashba R-1 compounds, i.e., in which the inversion symmetry is broken by dipoles generated by intrinsic polar atomic sites (steps 1 and 2 below). Such a database has not existed as yet and will be constructed below. After this we will down select those Rashba compounds that have band anti crossing (step 3 below). Our 3 steps are as follows:

(1) *Find nonmagnetic gapped compound calculated previously by DFTs* (filter 1 in Fig. 8): Our starting point is the aflow- ICSD- database (note that most of compounds in ICSD have been previously synthetized)[59], containing **20,831** unique compounds with less than 20 atoms per unit cell that were calculated by DFT (see Method section for details). Next, we down select (**filtering 1** in Fig. 8) those compounds that have time reversal symmetry (nonmagnetic) resulting in **13,838** non-magnetic compounds, from which **6355** are gapped nonmagnetic compounds (band gap larger that 1 meV). We note that the above-mentioned database used as magnetic configuration a ferromagnetic ordering.

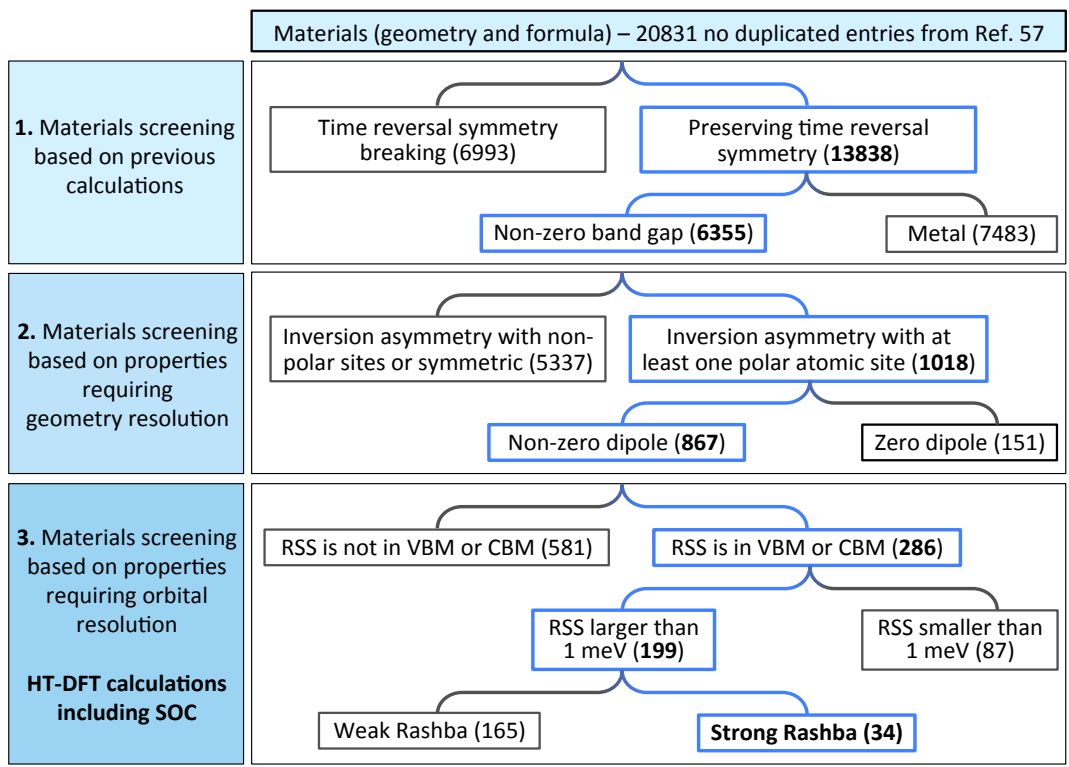

**Figure 8.** Schematic representations of filters applied to find strong Rashba material. These properties are separated in terms of the input required to compute them and also in terms of a binary selection that accept (blue) or reject (black) compounds.

(2*) Find the subset of non-magnetic gapped compounds that has non-centrosymmetric space group with at least one polar atomic site and non-zero dipole*, (**filter 2** in Fig. 8). We use the space group of the compounds to filter materials with polar atomic sites (polar space groups). The list of point groups with at least one polar site is given in Figure S1 of supplementary information IV. The cancelation of dipoles can be geometrical determined for each atomic site by considering vectors along the atomic bonds (details of the cancelation of dipole can be found in supplementary information IV). This gives **867** compounds that are Rashba nonmetals.

(3) S*ort out the subsets of Rashba nonmetals with anti-crossing bands (Type I Rashba) and with no anti crossing bands (Type II Rashba)*, (**filter 3** in Fig. 8): To do so we perform high-throughput DFT calculations including SOC of the band structure and spin-texture for the 867 Rashba nonmetals in order to identify anti-crossing bands and classify them into strong and weak Rashba compounds (DFT details are given in Methods). We find **286** Rashba compounds with spin splitting positioned within 30 meV or less from the band edges, the rest having spin split bands away from the band edges. Among these band-edge Rashba-insulators we find **199** have non-negligible spin splitting of 1 meV or more.

We next apply to these compounds our orbital projection analysis of band anti-crossing vs no band anti-crossing (See Fig. 4) to discern strong from weak effects. The distinction between anti-crossing and non-crossing bands (and hence between strong and weak Rashba effects) is evidenced by the change of the atomic orbitals weigh in the wavefunction around the momentum-offset $k_R$. Specifically, in non-crossing bands, the orbital character is essentially the same along all k-points (see Fig. 3b). However, for anti-crossing bands, the orbital character for k-vectors smaller and larger than the momentum-offset is expected to be different (see Fig. 3d). Additionally, the valence and conduction bands are made of different atomic orbitals (which are in different sites), as previously discussed. Using band anti-crossing, we identify for the final 199 selected Rashba compounds with spin splitting above 1 meV those having a strong Rashba effect. *This leads to **165 weak** and **34 strong** Rashba compounds that have been previously synthesized, most of them unappreciated as Rashba materials.*

## DISCUSSION
### Assessment of the predicted trends in strong Rashba compounds
We show that when the interaction between crossing bands is symmetry allowed, the induced anti-crossing leads to large Rashba spin splitting (strong Rashba compounds). We demonstrate that the *anti-crossing* is a design principle for the large Rashba coefficients in crystalline solids, in addition to the well established necessary but not sufficient conditions (NC space group, dipole generated by polar atomic sites and the presence of SOC). Notable trends include:

(i) An immediate consequence of the above noted design principle is that when topological insulators satisfy the symmetry condition to be Rashba compounds they must have a strong Rashba effect because TI intrinsically have band anti crossing. Because of limitations in the current listing of TI compounds[25–27,60] we find only two positive predictions of strong Topological Rashba Insulators (TlN and $Sb_2Se_2Te$).

Table II. Strong Rashba compounds with **Rashba spin splitting both in VBM and in CBM** and with large Rashba coefficient in at least one of these bands. For each compound, we present the ICSD code, space group, high symmetry $k$-point for the Rashba splitting in the valence ($K_v$) and conduction bands ($K_c$), Rashba spin splitting ($E_{Rv}$ and $E_{Rc}$) in meV, momentum offset ($k_{Rv}$ and $k_{Rc}$) in Å$^{-1}$, and the Rashba parameters ($\alpha_{Rv}$ and $\alpha_{Rc}$) in eVÅ.

| Material | ICSD | Space group | $K_v$ | $E_{Rv}$ | $k_{Rv}$ | $\alpha_{Rv}$ | $K_c$ | $E_{Rc}$ | $k_{Rc}$ | $\alpha_{Rc}$ |
|---|---|---|---|---|---|---|---|---|---|---|
| CsCuBi$_2$S$_4$ | 93370 | 36 | Γ | 41.5 | 0.258 | 0.322 | Y | 48.2 | 0.034 | 2.864 |
| SbF | 30411 | 40 | Y | 3.1 | 0.056 | 0.109 | Γ | 175.3 | 0.124 | 2.836 |
| KIO$_3$ | 97995 | 160 | Z | 17.4 | 0.055 | 0.628 | Z | 75.8 | 0.055 | 2.741 |
| PbS | 183249 | 28 | X | 1.7 | 0.035 | 0.099 | Y | 45.4 | 0.035 | 2.628 |
| ZnI$_2$O$_6$ | 54086 | 4 | Z | 16.8 | 0.334 | 0.101 | X | 111.0 | 0.091 | 2.448 |
| Ga$_2$PbO$_4$ | 80129 | 40 | R | 1.8 | 0.046 | 0.079 | Y | 144.4 | 0.119 | 2.428 |
| IrSbS | 74730 | 29 | U | 58.7 | 0.092 | 1.370 | Γ | 10.9 | 0.026 | 0.824 |
| Ga$_2$PbO$_4$ | 33533 | 1 | N | 1.1 | 0.020 | 0.116 | Y | 142.8 | 0.119 | 2.398 |
| CsPbF$_3$ | 93438 | 161 | Γ | 2.8 | 0.017 | 0.324 | Γ | 62.3 | 0.052 | 2.380 |
| PbS | 183250 | 28 | X | 1.2 | 0.034 | 0.071 | Y | 40.7 | 0.034 | 2.372 |
| KIO$_3$ | 247719 | 146 | Γ | 8.2 | 0.057 | 0.288 | Γ | 62.0 | 0.057 | 2.185 |
| KIO$_3$ | 424864 | 161 | Γ | 7.1 | 0.038 | 0.378 | Γ | 60.1 | 0.057 | 2.120 |
| PbTeO$_3$ | 61343 | 76 | X | 11.6 | 0.133 | 0.175 | M | 38.2 | 0.044 | 1.721 |

(ii) Based in our inverse design approach, we predict 13 previously synthesized but unnoticed as Rashba compounds with spin splitting in *both* VBM and CBM with large Rashba coefficient in at least one band (Table II), 9 Rashba compounds with spin splitting in *both* VBM and CBM with large Rashba coefficient in both bands (Table III), 9 compounds with strong Rashba effect in the CBM (Table IV), and additional 3 compounds with strong Rashba compounds in the VBM (Table V). The Rashba parameters, as well as the spin splitting and momentum offset are specified in Tables II-V. Band structures and spin texture of the predicted compounds are shown in supplementary information I. Compounds with the same symmetry, atomic identities, and composition can have different format of the Rashba bands or position of the band edges due to different temperatures or fabrication methods. We exclude compounds with very similar RSS or Rashba coefficient, listing then in Table II-V similar compounds with different RSS or band edges in different high symmetry $k$-points. These are also identified by different ICSD codes.

Table III. Strong Rashba compounds with **Rashba spin splitting both in VBM and in CBM**. For each compound, we present the ICSD code, space group, high symmetry $k$-point for the Rashba splitting in the valence ($K_v$) and conduction bands ($K_c$), Rashba spin splitting ($E_{Rv}$ and $E_{Rc}$) in meV, momentum offset ($k_{Rv}$ and $k_{Rc}$) in Å$^{-1}$, and the Rashba parameters ($\alpha_{Rv}$ and $\alpha_{Rc}$) in eVÅ.

| Material | ICSD | Space group | $K_v$ | $E_{Rv}$ | $k_{Rv}$ | $\alpha_{Rv}$ | $K_c$ | $E_{Rc}$ | $k_{Rc}$ | $\alpha_{Rc}$ |
|---|---|---|---|---|---|---|---|---|---|---|
| BiTeI | 74501 | 156 | A | 191.5 | 0.084 | 4.548 | A | 226.1 | 0.063 | 7.158 |
| BiTeI | 79364 | 156 | A | 187.9 | 0.084 | 4.475 | A | 218.8 | 0.063 | 6.948 |
| Sb$_2$TeSe$_2$ | 60963 | 160 | Γ | 101.6 | 0.052 | 3.885 | Γ | 144.0 | 0.045 | 6.402 |
| K$_2$BaCdSb$_2$ | 422272 | 26 | Γ | 18.6 | 0.016 | 2.356 | Γ | 41.5 | 0.016 | 5.251 |
| PbS | 183243 | 160 | L | 20.1 | 0.019 | 2.119 | L | 43.5 | 0.019 | 4.587 |
| BiTeCl | 79362 | 186 | Γ | 133.0 | 0.075 | 3.564 | Γ | 56.7 | 0.025 | 4.557 |
| GeTe | 659808 | 160 | L | 21.6 | 0.019 | 2.312 | L | 28.2 | 0.019 | 3.015 |
| GeTe | 188458 | 160 | Z | 142.5 | 0.068 | 4.219 | L | 25.0 | 0.019 | 2.686 |
| GeTe | 56040 | 160 | Z | 185.1 | 0.085 | 4.352 | L | 47.7 | 0.037 | 2.576 |

(iii) Considering the predicted compounds, we find cases that have a higher Rashba parameter than the largest currently known, e.g., 5.3 and 4.6 eVÅ for $K_2BaCdSb_2$ (*Pmc2₁*) and PbS (*R3m*), respectively. We also find giant RSS, even as larger as the previously reported for GeTe and BiTeI. For instance, for $Ga_2PbO_4$ (*Ama2*), the RSS is about 144 meV. Bands with the same representation are a required condition for large RSS in BiTeI[61], the anti-crossing reveals the relation of this condition with the orbital character and orbital interactions.

(iv) For direct band gaps compounds, we find that the Rashba splitting for the VBM and CBM occurs at the same TR-symmetry invariant *k*-point. In this case, the momentum offset is the same for both VBM and CBM, as predicted in our model, e.g., PbS (*R3m*), $KIO_3$ (*R3m*), $K_2BaCdSb_2$ (*Pmc2₁*), and $Sb_2Se_2Te$ (*R3m*). In general, the momentum offset is small, leading to large Rashba parameters even when the RSS is not large. On the other hand, compounds with indirect band gap can exhibit RSS at different *k*-points, i.e., the position of the VBM and CBM. This leads to *a)* compounds with RSS at different TR-symmetry points (Table II), and b) compounds with RSS at only one band edge (Tables III and IV). In this second group, the RSS is far from either the VBM or CMB; examples of this material include the KSnSb (*P6₃mc*) with RSS of 80 meV and Rashba parameter of 3.86 eVÅ in the CBM.

Table IV. Strong Rashba compounds with **Rashba spin splitting only in the CBM**. For each compound, we present the ICSD code, space group, high symmetry *k*-point for the Rashba splitting in the conduction bands ($K_c$), Rashba spin splitting ($E_{Rc}$) in meV, momentum offset ($k_{Rc}$) in Å$^{-1}$, and the Rashba parameters ($\alpha_{Rc}$) in eVÅ.

| Material | Space group | ICSD | $K_c$ | $E_{Rc}$ | $k_{Rc}$ | $\alpha_{Rc}$ |
|---|---|---|---|---|---|---|
| GeTe | 160 | 659811 | Z | 46.8 | 0.019 | 4.949 |
| KSnSb | 186 | 33933 | G | 80.2 | 0.042 | 3.862 |
| $Bi_2CO_5$ | 44 | 94740 | Z | 141.9 | 0.088 | 3.232 |
| KSnAs | 186 | 40815 | G | 39.2 | 0.025 | 3.079 |
| $TlIO_3$ | 160 | 62106 | Z | 56.9 | 0.052 | 2.184 |
| $Tl_3SbS_3$ | 160 | 603664 | Z | 89.9 | 0.083 | 2.169 |
| $CsGeI_3$ | 160 | 62559 | Z | 26.7 | 0.027 | 1.946 |
| AuCN | 183 | 165175 | L | 23.1 | 0.026 | 1.781 |
| $KCuBi_2S_4$ | 36 | 91297 | Y | 51.2 | 0.035 | 2.947 |

Table V. Strong Rashba compounds with **Rashba spin splitting only in the VBM**. For each compound, we present the ICSD code, space group, high symmetry *k*-point for the Rashba splitting in the conduction bands ($K_v$), Rashba spin splitting ($E_{Rv}$) in meV, momentum offset ($k_{Rv}$) in Å$^{-1}$, and the Rashba parameters ($\alpha_{Rv}$) in eVÅ.

| Material | Space group | ICSD | $K_v$ | $E_{Rv}$ | $k_{Rv}$ | $\alpha_{Rv}$ |
|---|---|---|---|---|---|---|
| $Te_7As_5I$ | 8 | 31877 | Z | 165.7 | 0.19 | 1.748 |
| LiSbZn | 186 | 642350 | G | 29.7 | 0.042 | 1.424 |
| LiSbZn | 186 | 42064 | G | 27.9 | 0.042 | 1.334 |

## Conclusion

In order to have a broad view of design principles for large Rashba parameters in solids, we perform large-scale-DFT calculations of more than 800 potential Rashba compounds. These calculations capture the physical mechanism determining the "Rashba scale", which is the based of the proposed theory here to explain and guide the selection of large Rashba compounds. Specifically, we show that when the interaction between crossing bands is symmetrically allowed, the induced anti-crossing leads to large Rashba spin splitting (strong Rashba compounds). We demonstrate that the *anti-crossing* is a design principle for the large Rashba coefficients in crystalline solids, in addition to the well established necessary but not sufficient conditions (NC space group and dipole generated by polar atomic sites). This establishes a causal relation between TIs and large Rashba coefficients, defining then the cross-functionality of TRIs. We used the proposed design principles as filters to distil from a large set of compounds those featuring strong Rashba effect. For instance, from lists of TIs, which intrinsically exhibit anti-crossing bands, filter compounds with the mentioned condition finding two positive predictions of strong Rashba compounds (TlN and $Sb_2Se_2Te$). In the same spirit, from the performed DFT calculations we filter compounds with anti-crossing bands, predicting 34 strong Rashba compounds, which included the known GeTe and BiTeI and the fabricated but unnoticed as Rashba compounds PbS (*R3m*), BiTeCl (*P6$_3$mc*), and BaCdK2Sb2 (*Pmc2$_1$*). These identified compounds provide a platform for spin-conversion devices and the exploration of phenomena potentially hosted by Rashba compounds.

## EXPERIMENTAL PROCEDURES
### Density functional calculations

The DFT band structure calculation were performed using the Perdew-Burke-Ernzenhof generalized gradient approximation (PBE)[64] exchange-correlation functional and the Hubbard on-site term[65,66] as implemented in the Vienna Ab-initio Simulation Package (VASP)[67,68]. We use the theoretical structures predicted in the AFLOW-database[59] by initially setting the magnetic configuration as ferromagnetic and non-magnetic and then performing the internal energy minimization of the experimental structure in the ICSD[43]. Our calculations were performed by assuming a non-magnetic configuration in the structures previously reported by Ref. [59] as non-magnetic. This could lead to some false positive non magnetic determinations as Ref. [59] decided if a structure is magnetic or not on the basis of a limited range of trial magnetic configurations (usually only FM) performed usually only with soft exchange correlation energy functional. All the specific settings of the calculations with spin-orbit coupling (e.g. cutoff energies, *k*-point sampling, effective U parameters) are the same as those used in Ref. [59].

High-throughput DFT-quantification of Rashba coefficients: For linear Rashba spin splitting, $\alpha_R$ is given by the ratio between the energy splitting and the momentum offset, i.e., $\alpha_R = 2E_R/k_R$. However, the value of the Rashba coefficient could depend on the symmetry path in the Brillouin zone (BZ)[20]. The orbital interaction, and hence the anti-crossing bands, can depends on the symmetry of the specific *k*-vector. However, we here report the Rashba coefficient at the VBM and CBM calculated along the symmetry directions connecting the high symmetry *k*-points. Additional analyses are required to study the specific conditions leading to anisotropic Rashba effect in each of the reported compounds. In this work, the Rashba coefficient for Rashba bands near the VBM or CBM is determined following these steps: *i*) we first identify TRIM points with spin splitting by looking at the energy difference of spin bands along the high symmetry path in the Brillouin zone; ii) performing the derivative of the energy dispersion with respect to the momentum, we select those TRIM points with changes in the sign of the derivatives for the upper (lower) band in the valence (conduction) band; iii) if the spin splitting is near the VBM (CBM) or less (more) than 30meV below (above) the VBM (CBM), we use the numerical value of the k-point in which the sign of the derivatives changes (i.e., the momentum offset $k_R$) and the value of the spin splitting ($E_R$) to compute the Rashba coefficient, i.e., $\alpha_R = 2E_R/k_R$. This procedure is performed in an automatic way for all DFT-calculated band structures.

**Orbital interaction in a one-dimensional model**

We here describe in more detail the proposed model for a one-dimensional chain of atoms, with two sites in the unit cell, one containing an *s*- orbital and other a *p*-orbital, as represented in Fig. 9. For simplicity, we consider that the 1D chain of atoms is along the *x*-axis, which imposes that the interaction between *s*- and $p_x$-orbitals is different from zero and the interaction between *s*- and $p_{yz}$-orbitals is symmetry forbidden. In the TB Hamiltonian, the matrix elements are given by

$$[H(k)]_{jj'}^{\sigma\sigma'} = \varepsilon_j \delta_{jj'} + \sum_v t_{jj'} e^{ik \cdot R_v},$$

where *j* and *σ* are the orbital (*s*- or $p_x$) and spin indexes (↑ or ↓), respectively. The considered hopping terms are the inter-site intra-orbital interaction (same orbital and same spin at different unit cells, i.e., $t_{ss}^{\sigma\sigma}$ and $t_{pp}^{\sigma\sigma}$), the on-site SOC (same orbital and different spin at different unit cells, i.e., $t_{ss}^{\uparrow\downarrow}$ and $t_{pp}^{\uparrow\downarrow}$), and the inter-atomic interaction ($t_{sp}^{\uparrow\downarrow}$) (See Fig. 9). The latter corresponds to the interaction between bands with different atomic-orbital characters, which we refer to hereinafter as band interaction. Thus, the Hamiltonian can be written as

$$H(k) = \begin{pmatrix} H_s(k) & H_{sp}(k) \\ H_{sp}^\dagger(k) & H_p(k) \end{pmatrix},$$

where local Hamiltonian $H_p(k)$ describing the interaction between *p*- orbitals is given by

$$H_p(k) = \begin{pmatrix} -\varepsilon_p + 2t_{pp}^{\uparrow\uparrow}\cos(k_x a) & -i2t_{soc}^p \sin(k_x a) \\ i2t_{soc}^p \sin(k_x a) & -\varepsilon_p + 2t_{pp}^{\downarrow\downarrow}\cos(k_x a) \end{pmatrix}.$$

The breaking of the inversion symmetry is introduced by imposing that the SOC (i.e., the interaction between different spins $t_{pp}^{\uparrow\downarrow}$) satisfy the relation $t_{pp}^{\uparrow\downarrow}(r) \neq t_{pp}^{\uparrow\downarrow}(-r)$. Specifically, we consider that $t_{pp}^{\uparrow\downarrow}(r) = -t_{pp}^{\uparrow\downarrow}(-r) = -t_{soc}^p$. As shown in the above Hamiltonian $H_p(k)$, this approximation gives the off-diagonal matrix element $[H_p(k)]_{pp}^{\uparrow\downarrow} = t_{pp}^{\uparrow\downarrow}(a)e^{ika} + t_{pp}^{\uparrow\downarrow}(-a)e^{-ika} = -i2t_{soc}^p \sin(k_x a)$. This symmetry based approximation leads to the same results expected in a $k \cdot p$ model (e.g., the Hamiltonian in Eq. 1) using the $L \cdot S$ term (i.e., the Rashba term $\alpha_R \sigma_y k_x$ in one-dimensional system), as we show below.

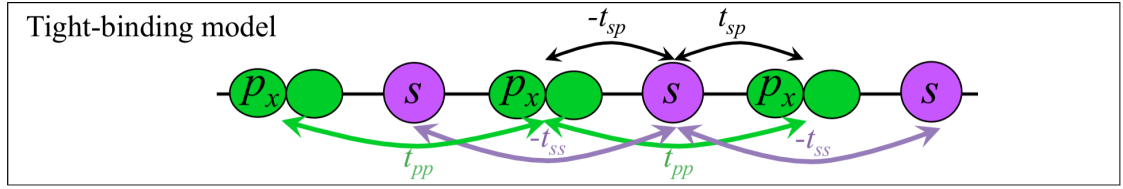

**Figure 9.** The TB model: One dimension chain formed by s (magenta) and $p_x$ (green) orbitals in different sites. The inter-orbital interaction $t_{sp}$ and intra-orbital interactions $t_{ss}$ and $t_{pp}$ are schematically defined.

For $k \to 0$, considering that $t_{pp} = t_{pp}^{\uparrow\uparrow} = t_{pp}^{\downarrow\downarrow}$, this Hamiltonian results in a very simplified expression for the p- orbital interaction, namely,

$$H_p(k) \approx \begin{pmatrix} -\varepsilon_p + 2t_{pp} - t_{pp}a^2 k_x^2 & -i2t_{soc}^p k_x a \\ i2t_{soc}^p k_x a & -\varepsilon_p + 2t_{pp} - t_{pp}a^2 k_x^2 \end{pmatrix}.$$

This expression can be rewritten as $H_p(k) \approx \sigma_0(-\varepsilon_p + 2t_{pp} - t_{pp}a^2 k_x^2) + 2at_{soc}^p(\sigma_y k_x)$. In quasi two-dimensional compounds, the SOC gives an equivalent expression for the off-diagonal matrix elements, i.e., $2at_{soc}^p(\sigma_y k_x - \sigma_x k_y)$. This reproduces phenomenological Hamiltonian in Eq. 1, which intrinsically leads to the weak Rashba effect (i.e., small Rashba coefficient). From the off-diagonal term $2at_{soc}^p(\sigma_y k_x - \sigma_x k_y)$, the Rashba parameter can easily identified as $\alpha_R = 2at_{soc}$. Here, the eigenvalues of $H_p(k)$ are given by $E_{p\pm}(k) = -\varepsilon_p + 2t_{pp} - t_{pp}a^2 k_x^2 \pm \alpha_R |k_x|$, which intrinsically accounts for the Rashba spin splitting in weak Rashba compounds as given by Eq. 1. According to our results, Eq. 1 can describe the spin splitting in weak Rashba compounds, even in 3D materials (as shown $KSn_2Se_4$). The dependence of the Bulk Rashba effect in 3D compounds with respect to the interatomic orbital interaction is essentially given by the energy band anti-crossing, which is the metric defining the Rashba scale.

Analogously for interactions only between p-orbitals, we have $H_s(k) \approx \sigma_0(\varepsilon_s - 2t_{ss} + t_{ss}a^2 k_x^2) - 2at_{soc}^s(\sigma_y k_x)$. Finally, the matrix $H_{sp}(k)$, without loss of generality, counts for the interaction between s- and p- orbitals with different spin, i.e.,

$$H_{sp}(k) = \begin{pmatrix} 0 & -i2t_{sp}^{\uparrow\downarrow}\sin(k_x a) \\ i2t_{sp}^{\uparrow\downarrow}\sin(k_x a) & 0 \end{pmatrix}.$$

For weak inter-orbital interaction $t_{sp}^{\uparrow\downarrow}$ is smaller. In that case, the Hamiltonian $H(k)$ can approximately be treated as block diagonal, where blocks separately describe s and p orbitals, and hence, the Rashba parameter is approximately given by $\alpha_R = 2at_{soc}$.

The crossing between bands meanly formed by s- and p-orbitals only depends on the relative on-site energy between orbitals $\Delta_{sp} = (\varepsilon_s - 2t_{ss}) - (\varepsilon_p - 2t_{pp})$ and the intra-orbital interaction $t_{ss}$ and $t_{pp}$. For instance, for $t_{ss} = t_{pp}$, bands cross when $\Delta_{sp} < 0$. The role of the inter-orbital interaction $t_{sp}^{\uparrow\downarrow}$ is to open the band gap, which increase as this interaction increase. For this reason, strong Rashba semiconductors (compounds exhibiting anti-crossing) usually also have smaller band gaps. In general, band anti-crossing can be designed in a periodic Hamiltonian by requiring a non-zero interaction between at least two different atomic orbitals with opposite effective mass, as illustrated in the proposed one-dimensional chain with two atomic species.

Note that we used here the *s*- orbitals as notation for states with total angular momentum equal to J=1/2, and hence, the discussion previously presented is for instance also extended to *p$_z$*-orbitals, which leads to a non-zero SOC. The pure s- orbitals should results in a zero Rashba spin splitting since the SOC is zero. In fact, the obtained Hamiltonian $H(k)$ is similar to that discussed in Ref. [12] for the interaction between states with total angular momentum *J*=1/2 and *J*=3/2.

### DATA AND SOFTWARE AVAILABILITY
All data needed to evaluate the conclusions in the paper are present in the paper and the Supplemental Information. Additional data related to this paper may be requested from the authors.

### Acknowledgment

The work at the University of Colorado at Boulder was supported by the National Science Foundation NSF Grant NSF-DMR-CMMT No. DMR-1724791. The work in Brazil was supported by the Sao Paulo Research Foundation FAPESP Grant No. 17/02317-2 and by CNPq. CMA and EO were supported by the Sao Paulo Research foundation FAPESP grants 18/11856-7 and 18/11641-0, respectively. The high-throughput first-principle calculations were performed using the computational infrastructure of the LNCC supercomputer center (Santos Dumont) in Brazil. Big-data analysis was performed using the computational infrastructure of USA NSF XESED.

### AUTHOR CONTRIBUTIONS
All authors participated in the conceptual development of the project. C.M.A. and E.O performed the calculations. C.M.A. and A.Z made the analysis of the results and wrote the paper with input from all authors. A.Z, AF, and G.M.D. directed the study.

### DECLARATION OF INTEREST
The authors declare no conflict of interest.